\documentstyle[12pt,fleqn]{article}
\def\soc{{\rm C}_{60}}
\def\rug{{\rm C}_{70}}
\def\c76{{\rm C}_{76}}
\def\c78{{\rm C}_{78}}

\def\beeq{\begin{equation}}
\def\eneq{\end{equation}}
\def\beeqa{\begin{eqnarray}}
\def\eneqa{\end{eqnarray}}

\addtocounter{section}{-1}
\begin{document}

\begin{center}

\vspace{2in}

{\large {\bf{Optical absorption spectra
of $A_{\bf 6}$C$_{\bf 60}$ and $A_{\bf 6}$C$_{\bf 70}$:\\
Reduction of effective Coulomb interactions\\
in Frenkel excitons
} } }

\vspace{1cm}

{\rm Kikuo Harigaya}\\

\vspace{1cm}

{\sl Fundamental Physics Section,\\
Electrotechnical Laboratory,\\
Umezono 1-1-4, Tsukuba, Ibaraki 305, Japan}

\vspace{1cm}

(Received~~~~~~~~~~~~~~~~~~~~~~~~~~~~~~~~~~~)
\end{center}

\Roman{table}

\vspace{1cm}

\noindent
{\bf ABSTRACT}\\
We theoretically investigate optical absorption spectra of $\soc^{6-}$
and $\rug^{6-}$, and discuss relations with the optical properties of
alkali metal doped fullerides $A_6\soc$ and $A_6\rug$.  This is a
valid approach for systems where Frenkel exciton effects are dominant.
We use a tight binding model with long ranged Coulomb interactions and
bond disorder.  Optical spectra are obtained by the Hartree-Fock
approximation and the configuration interaction method.  We find that
the Coulomb interaction parameters, which are relevant to the optical
spectra of $A_6\soc$ ($A_6\rug$) in order to explain the excitation
energies and relative oscillator strengths of absorption peaks, are
almost the half of those of the neutral $\soc$ ($\rug$).  The reduction
of the effective Coulomb interactions is concluded for the heavily doped
case of $\soc$ and $\rug$.  This finding is closely related with the
experimental fact that dielectric constants of fullerides which
are maximumly doped with alkali metals become about twice as large as
those of the neutral systems.

\mbox{}

\noindent
PACS numbers: 78.66.Qn, 78.20.Dj, 71.35.+z, 31.20.Tz

\pagebreak

\section{INTRODUCTION}

Since the $\soc$ solid and $A_3\soc$$^{1}$ with the novel high
temperature superconductivity were discovered, fullerenes have
been intensively investigated.  As the $\pi$ electrons are
delocalized on their surfaces, the fullerenes show optical
responses that are similar to those in $\pi$ conjugated polymers.$^{2}$
For example, the absorption spectra of $\soc$ (Refs. 3 and 4)
and $\rug$ (Ref. 3) reflect the existence of excitons (mainly
Frenkel excitons) which are important when the excitation
energy is larger than the order of 1eV.  The nonlinearity of
$\soc$ in the third harmonic generation (THG) is of the order
$10^{-11}$esu,$^{5,6}$ and the similar magnitudes have been
observed in polydiacetylenes.

Recently, we have studied the linear absorption and the THG of
$\soc$ by using a tight binding model$^{7}$ and a model with
a long ranged Coulomb interaction.$^{8,9}$  A free electron
model yields the THG magnitudes which are in agreement with the
experiments of $\soc$.$^{5,6}$  However, when the Coulomb interactions
are taken into account, the THG magnitudes decrease.$^{8}$  We
have discussed that the local field correction would be necessary
in order to recover the agreement.  The model with Coulomb interactions
has turned out to describe well the linear absorption spectra of
$\soc$ and $\rug$ in solutions.$^{9}$

In the first half part of this paper, we perform calculations for
$\soc^{6-}$ and consider the optical spectra of $A_6\soc$
($A =$ K, Rb, Cs, etc).  This compound is a fulleride maximumly
doped with alkali metals, and is an insulator like the neutral $\soc$.
The optical spectra have been measured in several papers.$^{10,11}$
The peak structures in the energy dependence are quite different from
those of the neutral $\soc$, even though each peak could be explained
as an optical transition between molecular orbitals.  The absorption
spectra of the neutral $\soc$ and $A_3\soc$ are rather similar, but
the data of $A_6\soc$ are largely different.  It seems that main peaks
of the neutral $\soc$ spectra move to lower energies.  This fact could
be explained by reduction of the effective Coulomb interaction strengths.
The principal purpose of the present calculations is to confirm this
viewpoint.  The calculation method is the same as that used in the
previous paper.$^9$  We start from the Hartree-Fock approximation and
perform configuration interaction calculations which are limited to
single electron-hole excitations (single CI).  In the experiments
of $A_6\soc$,$^{10,11}$ intermolecular interactions seem to
be relatively weak, because there is not a peak structure owing
to the intermolecular interactions (like the 2.8eV structure in
the $\soc$ films$^{4}$).  Therefore, we shall limit our
considerations to the system $\soc^{6-}$.  This is a reasonable
approximation for systems where single molecular contributions
(Frenkel exciton effects) are dominant.

The reduction of the Coulomb interactions means that dielectric
constants of the doped systems increase.  Actually, the
electron-energy-loss spectroscopy (EELS) studies of Rb$_x \soc$
and Rb$_x \rug$ ($x = 0, 3, 6$)$^{12}$ have revealed that the real part
of the dielectric function at zero frequency increases from
$x=0$ to 6, and the magnitude becomes about twice.  In our calculations
by single molecule models, this enhancement of the dielectric
functions is represented by the decrease of the phenomenological
parameters in Coulomb interactions.

In the last part of the paper, we consider $\rug^{6-}$ in order
to discuss optical properties of the insulating phase $A_6\rug$.
Frenkel excitons would be dominant in this system again.
We demonstrate that the picture of the reduction of Coulomb
strengths is relevant to $\rug$ also.  Therefore, the reduction
of Coulomb interaction strengths is a common property seen in
two kinds of fullerenes.  In fact, the EELS studies$^{12}$
have concluded that $\pi$ electronic systems in Rb$_x\soc$
and Rb$_x\rug$ show a non-rigid-band behavior and excitation
energies become smaller upon doping.

In the next section, we explain our model briefly.  In Sec. III,
we show results of $\soc^{6-}$ and give discussion relating
with molecular orbital structures. In Sec. IV, we consider $\rug^{6-}$.
The final section is devoted to the summary.

\noindent
{\bf omitted afterwards ...}

\end{document}